\documentclass[aps,amsmath,twocolumn,amssymb,pre,showpacs]{revtex4}
\usepackage{amsmath,amssymb}
\usepackage{graphicx}
\usepackage{epsfig}
\usepackage{color}

\bibstyle{apsrev.bib}
\newcommand{\be}{\begin{equation}}
\newcommand{\ee}{\end{equation}}
\newcommand{\beqn}{\begin{eqnarray}}
\newcommand{\eeqn}{\end{eqnarray}}

\begin{document}
\input epsf.sty

\title{Disordered contact process with asymmetric spreading}

\author{R\'obert Juh\'asz}
\affiliation{Institute for Solid State Physics and Optics, Wigner Research Centre for Physics, H-1525 Budapest, P.O. Box 49, Hungary}
\email{juhasz.robert@wigner.mta.hu}
\date{\today}

\begin{abstract}
An asymmetric variant of the contact process where the activity spreads 
with different and independent random rates to the left and
to the right is introduced. A real space renormalization
scheme is formulated for the model by means of which
it is shown that the local asymmetry of spreading is irrelevant on large
scales if the model is globally (statistically) symmetric.
Otherwise, in the presence of a global bias in either direction, 
the renormalization method predicts two distinct 
phase transitions, which are related to the spreading of activity in and
against the direction of the bias. 
The latter is found to be described by an
infinite randomness fixed point while the former is not.    
\end{abstract}


\maketitle

{\it This paper is dedicated to Ferenc Igl\'oi on the occasion of his 60th birthday.}

\section{Introduction}
The contact process \cite{cp,liggett} is a simple of model of an epidemic that
is defined on a lattice of binary variables which can be either
inactive (healthy) or active (infected). 
The dynamics consist of two kinds of competing moves: infection of adjacent healthy sites by
infected ones and spontaneous healing of the latter. 
Besides being a starting point for designing more realistic models of
epidemics it is a thoroughly studied (but non-soluble), paradigmatic model 
of systems undergoing a non-equilibrium phase transition 
from a fluctuating phase to an absorbing one \cite{md,odor,hhl}.
In the case of translational invariance, the phase transition falls into
the universality class of directed percolation \cite{odor}. 
This type of critical behavior is, however, rarely observed in real systems \cite{takeuchi}, 
which is attributed to the fact that speading processes 
usually take place in inhomogeneous environments \cite{hinrichsen}. 
Indeed, according to the Harris criterion, the critical point of the pure
system is unstable against weak quenched disorder in dimensions $d<4$
\cite{noest}, and, here, 
the field-theoretic renormalization of the problem has 
only runaway solutions \cite{janssen}. 
The understanding of the behavior of the model in the presence of quenched
disorder is therefore of great importance. 
Early Monte Carlo simulations showed, in agreement with phenomenological
considerations, that the disorder leads to anomalous dynamics outside 
of the critical point \cite{noest,moreira,cgm,szabo} 
and, at the critical point, the different observables scale with 
some power of the logarithm of time rather than the 
time itself \cite{moreira}. 
Below the critical point, namely, the density of active sites and 
the survival probability show a power-law decay with dynamical 
exponents varying continuously
with the control parameter, while, above the critical point, the size of 
the set of active sites is growing
sub-linearly in one dimension when the process was started 
from a single active seed \cite{bramson}.   
These phases are analogues of the disordered and ordered Griffiths phases of
magnetic systems \cite{Griffiths}, respectively, 
and the slow dynamics are related to the
occurrence of so called rare regions which are locally in the opposite phase
with respect to the majority of the system. 
In the sub-critical Griffiths phase, the rare regions are locally
super-critical clusters and, 
although, being exponentially rare (in their size), due to
their exponentially large extinction time they are able to change the usual
exponential temporal decay of the density to an algebraic one.
Instead, in the super-critical Griffiths phase, the creeping motion of the
front of the active cluster is caused by rare sub-critical regions which impede the spreading of activity for long times.    
A substantial progress in the quantitative description of the critical point
was the adaptation of a strong disorder renormalization group (SDRG) scheme
\cite{im} to the model by Hooyberghs, Igl\'oi and Vanderzande \cite{hiv}.
The formal description of the model by the SDRG is essentially 
identical with that of the random transverse-field Ising model \cite{fisher}.
For sufficiently strong disorder, the critical behavior of the one-dimensional
model has been found to be controlled by an 
{\it infinite randomness fixed point}, 
where the dynamics are logarithmically slow and the critical
exponents are universal, i.e. independent of the form of disorder \cite{hiv}.
Later, large-scale Monte Carlo simulations have demonstrated that 
the predictions of the SDRG method are valid even for relatively 
weak disorder \cite{vd}.  
Nevertheless, the question whether the infinite-randomness fixed point 
is attractive for any weak disorder or
there exists a line of disorder-dependent fixed points for 
weak disorder is not settled yet \cite{hiv,nft,hoyos}.

In the SDRG treatment of the model, the activity is assumed to spread 
{\it through a given link} in both directions with the same rate (that varies
from link to link), while in the numerical simulations, like in Ref. \cite{vd}, slightly differently,
the infection {\it from a given site} occurs 
in both directions with equal rates. 
So, in the latter case, the spreading of activity through a link is
non-symmetric (or biased) and one may ask whether this case can be
treated by an appropriately generalized SDRG scheme.
More generally, one could pose the same question as well as the question 
what is the nature of 
the phase transition in an asymmetric variant of the disordered contact
process where the infection rates $\lambda_{ij}$ assigned to a directed link 
$(i,j)$ are completely uncorrelated random variables. 
In the case of the translationally invariant model with a bias (i.e. different
infection rates in the two directions) \cite{schonmann}
the phase transition falls into the directed percolation universality class as in the unbiased model \cite{tretyakov}. 
But, the breaking of the 
translation symmetry, for instance, by putting an active wall to the system,
leads to differences compared to the unbiased case, such as the discontinuity
of the velocity of the activity front across the transition \cite{costa}.    
So, one expects that in the simultaneous presence of a global bias and
quenched disorder the nature of the phase transition is 
different from that of the homogeneous, biased contact process. 
Another interesting feature of the homogeneous biased model is that, 
besides the absorbing phase transition, a second
(higher) transition point can be defined above (below) which the probability
that the origin is active as $t\to\infty$ is positive (zero) \cite{2nd}. 

The interplay between heterogeneity and bias
has also been studied in a similar model, 
the symmetric contact process with an additional
drift of the activity \cite{nelson,lebowitz,kessler}. 
This was motivated by the description of the
population dynamics of bacterial colonies in the presence of a flow in a
medium with a heterogeneous distribution of nutrients. 
In the continuum mean-field limit of this model in a random environment, a
delocalization transition has been found which is controlled by the convection
velocity \cite{nelson}. 
  
The aim of the present work is to study the disordered, asymmetric 
contact process in one dimension by working out and applying an SDRG scheme. 
We will show that, if the model is statistically symmetric, then the
local asymmetry of links is irrelevant and the critical behavior is identical
to that of the symmetric model. If, however, the model is globally asymmetric,
i.e. there is a bias in either direction, two distinct phase transitions
arise, which are related to the spreading of activity in and against the
direction of the bias. 
These results will be expounded in the rest of the paper, 
which is organized as follows. 
In Section \ref{model}, the precise definition of the model is 
given and the SDRG
scheme is introduced. The case of weak asymmetry is analyzed in Section
\ref{weak}, while the behavior of the globally symmetric and the biased model
is discussed in Section \ref{symmetric} and Section \ref{biased}, 
respectively. The results are summarized in
Section \ref{summary} and some calculations are presented in the Appendix.

\section{The asymmetric contact process and the SDRG scheme}   
\label{model}

Let us consider a one-dimensional
lattice where each site can be in two states: either active (infected) or
inactive (healthy). 
The asymmetric contact process is a continuous-time Markov process 
on this state space, which consists of the following transitions 
occurring independently. 
Active sites become inactive with a rate $\mu_i$ on site $i$, and, 
if site $i$ is active, it infects the neighboring site on its right with
a rate $\lambda_i$ and the neighboring site on its left with
a rate $\kappa_{i-1}$. The rates $\lambda_i$ and $\kappa_i$ will be termed as
forward and backward infection rates, respectively.  
The transition rates $\{\mu_i,\lambda_i,\kappa_i\}$ are time-independent random
variables, the distribution of which will be specified later.  

The SDRG transformation of the disordered, 
symmetric contact process\cite{hiv} is a 
real space renormalization transformation 
which sequentially eliminates the quickly relaxing 
degrees of freedom and replaces the original system by a reduced one 
that, however, preserves the slowly relaxing degrees of freedom of the
original system. 
Formally, finite blocks of sites are considered, 
the time evolution of which are governed by a master equation
\be 
\partial_t{\bf P}={\bf P}Q,
\ee
where the probabilities of states are arranged in the row vector 
${\bf P}(t)=(p_1(t),p_2(t),\dots)$ and $Q$ is the rate matrix (or infinitesimal
generator) of the process. 
In the SDRG procedure, the blocks are selected in which the spectrum of $Q$
shows a separation of levels. 
Then the higher-lying states are discarded and the block is replaced 
by a simpler one having the same low-lying levels. 
For the symmetric model, there are two types of reduction steps, and this
structure of the SDRG with appropriate modifications 
will be kept also for the asymmetric contact process. 

\subsection{Cluster merging}  

\begin{figure}
\includegraphics[width=2cm]{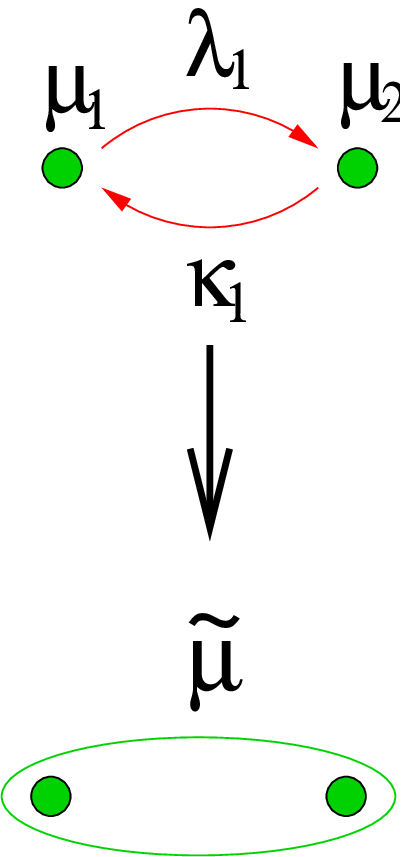}
\includegraphics[width=3.5cm]{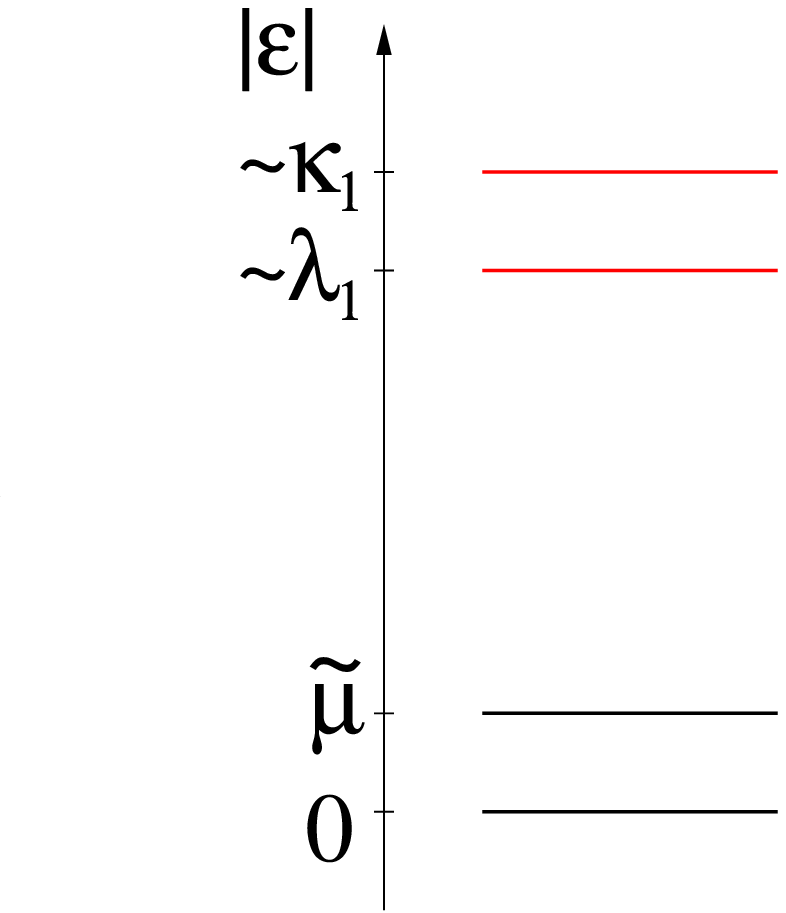}
\caption{\label{rg} (Color online) Left: Illustration of the merging of two sites connected
  by a link with large infection rates. 
Right: Schematic spectrum of the rate matrix of the two-site block.}
\end{figure}
First, let us consider an isolated block of two sites shown in Fig. \ref{rg}, 
and assume that the infection
rates in both directions are much larger than the recovery rates, 
$\lambda_1,\kappa_1\gg \mu_1,\mu_2$. 
In this case the two sites are most of the time either both active or both
inactive. This suggests that the block can be substituted by a single
giant site (or cluster) with an appropriate recovery rate $\tilde \mu$. 
Indeed, analyzing the spectrum of the rate matrix of the block, 
see the Appendix, one obtains that the two higher-lying  
levels are well separated from the two lower-lying ones. 
Note that at least one of the eigenvalues is always zero since the rate
matrix is a stochastic matrix.  
For an isolated single site with recovery rate $\tilde\mu$, the only 
non-trivial eigenvalue is $-\tilde\mu$. 
So, the effective recovery rate of the giant site is identified with the
magnitude of lowest
non-trivial eigenvalue of the rate matrix or, equivalently the (in magnitude)
smallest root of Eq. \ref{cubic}. 
If $\lambda_1,\kappa_1\gg \mu_1,\mu_2$, one obtains the approximate form: 
\be 
\tilde\mu\simeq\frac{\mu_1\mu_2}{\omega_1},
\label{mu_appr}
\ee
where the combined infection rate $\omega_i$ is defined as 
\be
\omega_i\equiv\frac{\lambda_i\kappa_i}{\lambda_i+\kappa_i}.
\ee   
Note that if the infection rates $\lambda_i$ and $\kappa_i$ differ by
orders of magnitudes then $\omega_i$ gives roughly the smaller one of them.  
We can see that this type of reduction of the two-site block to a single site 
is justified only if
{\it both} infection rates on the link (or, equivalently $\omega_i$) 
are much larger than the recovery
rates \footnote{This situation is somewhat reminiscent of the difficulties with the renormalization of the
  random XY chain where both couplings $J^x$ and $J^y$ should be larger than
  the other couplings \cite{fisher_xy}.}, 
and the life-time of activity in the two-site block is controlled 
by the smallest
infection rate. Consequently, in a totally asymmetric contact process where
infection spreads unidirectionally, giant clusters with low decay rate never
form, hence no rare region effects are expected to emerge 
(unless the initial recovery rates are not bounded away from zero).   

Besides the transition rates, we also keep track of the mass of clusters, 
i.e. the number $n$ of original sites comprised by a giant site. 
Obviously, this quantity transforms as 
\be
\tilde n=n_1+n_2.
\ee

\subsection{Cluster elimination}

\begin{figure}
\includegraphics[width=3cm]{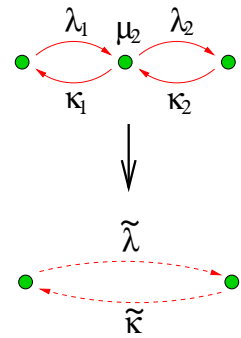}
\caption{\label{rg2} (Color online) Illustration of the elimination of a site with a large
  recovery rate $\mu_2$.}
\end{figure}

Next, let us consider an isolated three-site block shown in Fig. \ref{rg2}, where the recovery rates on the two lateral sites are zero. 
If the recovery rate $\mu_2$ is much larger than the infection rates,\be 
\mu_2\gg\lambda_1,\kappa_1,\lambda_2,\kappa_2,
\label{cond1}
\ee
the middle site will be most
of the time inactive and transmits infection only during its rare flashes of
activity. 
It is thus reasonable to eliminate this site and connect the lateral
sites by a direct link with some effective (reduced) infection rates. 
Solving the eigenvalue problem of the rate matrix of the block, see the
Appendix and Fig. \ref{spectrum}, one sees that the higher-lying $4$ levels, 
being in the order of
$\mu_2$, are well separated from the lower-lying $4$ levels among which 
two levels have zero eigenvalue. 
\begin{figure}
\includegraphics[height=4cm]{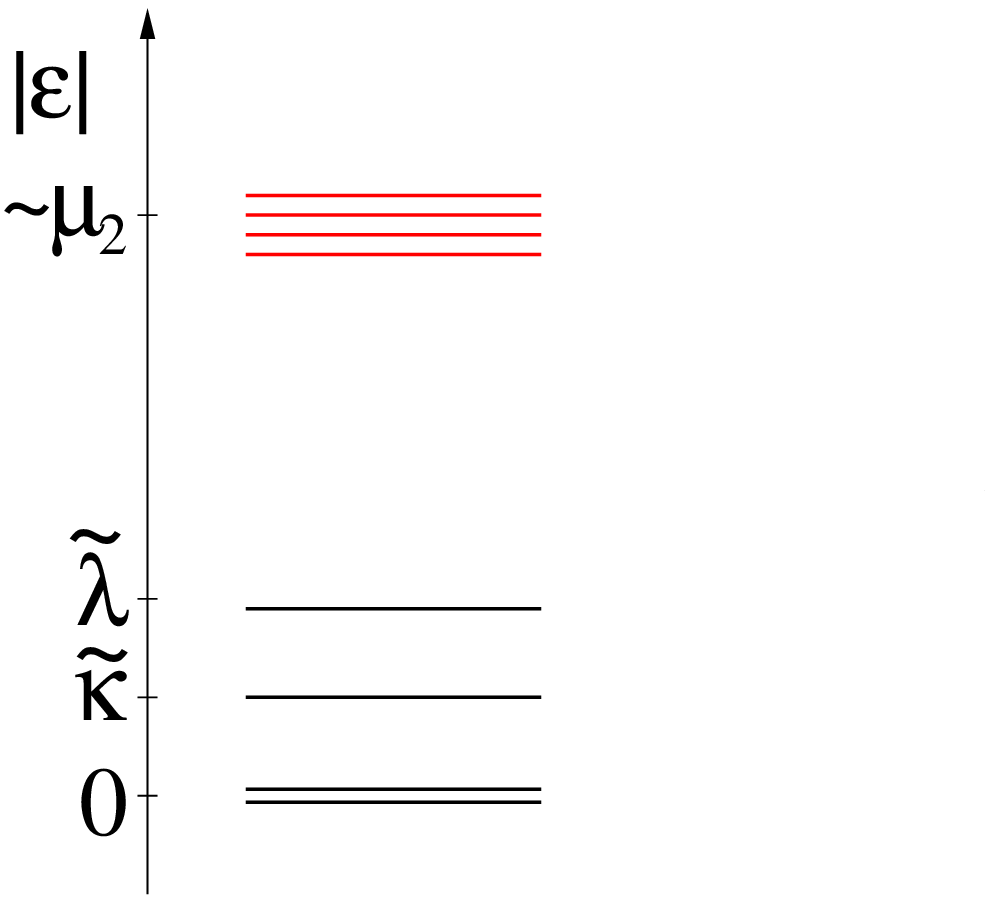}
\includegraphics[height=4cm]{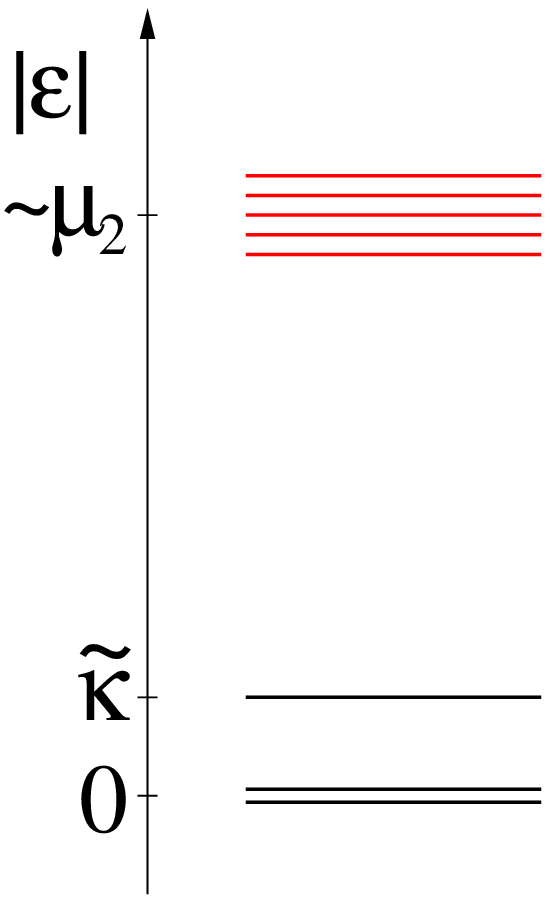}
\caption{\label{spectrum} (Color online) Schematic spectrum of the rate matrix of a
  three-site block in the case of complete separation (left) and partial separation (right).}
\end{figure}
This lower-lying bunch of levels is to be
compared with the spectrum of a two-site block with infection rates
$\tilde\lambda,\tilde\kappa$ and zero recovery rates. 
The eigenvalues of the latter are $(0,0,-\tilde\lambda,-\tilde\kappa)$, so the 
effective infection rates of the three-site block are identified as the
magnitude of the two lowest-lying non-trivial eigenvalues. 
If the condition in Eq. (\ref{cond1}) is fulfilled, we have
\beqn 
\tilde\lambda=s_{\lambda}
\left(1-\sqrt{1-\frac{\lambda_1\lambda_2}{s_{\lambda}^2}}\right), \quad s_{\lambda}\equiv\frac{\mu_2+\lambda_1+\lambda_2}{2} \label{lambdaeff} \\
\tilde\kappa=s_{\kappa}\left(1-\sqrt{1-\frac{\kappa_1\kappa_2}{s_{\kappa}^2}}\right),
\quad s_{\kappa}\equiv\frac{\mu_2+\kappa_1+\kappa_2}{2}
\label{kappaeff} 
\eeqn
and obtain the approximate expressions 
\be 
\tilde\lambda\simeq\frac{\lambda_1\lambda_2}{\mu_2},
\qquad 
\tilde\kappa\simeq\frac{\kappa_1\kappa_2}{\mu_2},  
\label{inf_appr}
\ee
while the other non-trivial eigenvalues are $O(\mu_2)$. 
Here, we speak of a {\it complete separation} of levels.
As we will see later, the cluster
elimination step will be applied in the SDRG procedure only 
if $\mu_2>\omega_1,\omega_2$ and, due to the broad distribution of
rates we have typically $\omega_i\approx\min \{\lambda_i,\kappa_i\}$.
If the two links of the block are biased in opposite directions, e.g. 
$\lambda_1,\kappa_2\ll\mu_2$ and  $\kappa_1,\lambda_2>\mu_2$,
then the spectrum still displays a weaker form of complete separation
\footnote{
To have a weaker form of complete separation it is sufficient to fulfill
the conditions $\min\{\lambda_1,\lambda_2\}\ll \mu_2$ and
$\min\{\kappa_1,\kappa_2\}\ll \mu_2$. Then the lowest-lying (non-trivial) 
eigenvalues are $O(\min\{\lambda_1,\lambda_2\})$ and
$O(\min\{\kappa_1,\kappa_2\})$, while the other eigenvalues are $O(\mu_2)$.}. 
But if the two links are biased in the same direction, e.g. to the right,  
\be
\lambda_1,\lambda_2>\mu_2, \quad \kappa_1,\kappa_2\ll\mu_2,
\ee
then only the lowest (non-trivial) eigenvalue separates from the higher
ones. 
The corresponding rate $\tilde\kappa\simeq\kappa_1\kappa_2/\mu_2$ can
still be interpreted as an effective spreading rate of the activity 
from right to left but by this
reduction, the information on how the activity spreads from left
to right is lost.   
We call this situation a {\it partial separation} of levels.

We will also keep track of evolution of 
the number $l$ of original links which are
incorporated into the effective links. This quantity is dual to the 
mass of giant sites and transforms in the cluster elimination step as 
\be
\tilde l=l_1+l_2.
\label{l}
\ee 

\subsection{The SDRG scheme}

After introducing the elementary reduction steps, we are in a
position to formulate a SDRG scheme for the asymmetric contact process. 
First, the largest one among the rates $\{\omega_i,\mu_i\}$, which will be
denoted by $\Omega$, is selected. 
If $\Omega$ is a combined infection rate (recovery rate) 
then the cluster merging step (cluster elimination step)
is performed. 
The above steps are then iteratively applied to the renormalized system.  
This recursive procedure gradually reduces the cut-off rate $\Omega$, which
amounts to that, roughly speaking, the fast processes occurring on a time scale
shorter than $\Omega^{-1}$ are eliminated. 
The basic dynamical relation between the time and the length scale 
can be inferred from the asymptotic dependence of 
the cut-off $\Omega$ on the fraction $f$ of active (non-decimated) degrees of freedom \cite{im}. 
Note that, if the model is symmetric ($\lambda_i=\kappa_i$) the 
above SDRG procedure reduces to the that applied in Ref. \cite{hiv}.

Before we proceed with the analysis of the SDRG transformation, 
we need to recapitulate some of the results of the symmetric contact process
obtained by the SDRG technique. 
For further details we refer the reader to Ref. \cite{hiv}. 
In the case of the symmetric model, the distribution of the logarithm of rates
is broadening without limits under the SDRG transformation, therefore 
the conditions of the approximative reduction steps are fulfilled
better and better and the procedure becomes asymptotically exact.
Due to that the rates at different sites remain independent provided they were
so initially, it is sufficient to deal with the evolution of the probability
of rates under the renormalization and the problem is 
analytically tractable.  
The asymptotic transformation of rates in Eqs. (\ref{mu_appr}) 
and (\ref{inf_appr}) is identical to that of the couplings and external fields of the random transverse-field Ising chain up to a constant factor which is irrelevant at the critical point \cite{hiv}. 
The asymptotic solution of the renormalization equations 
of the latter problem at the
critical point, which is at the self-dual point of the transformation, 
have been found in Ref. \cite{fisher}.    
Here, in the so called infinite randomness fixed point 
of the transformation, 
both the logarithm of infection and recovery rates scale 
with $\xi\equiv f^{-1}$ as  
\be
|\ln\tilde\lambda|\sim|\ln\tilde\mu|\sim\xi^{1/2}
\label{symm_crit}
\ee 
and the mass of clusters as well as the length of effective links grow as 
\be
\tilde n\sim\tilde l\sim\xi^{d_f}, 
\ee
where $d_f=(1+\sqrt{5})/4=0.809\cdots$ is the fractal dimension of clusters. 
The Griffiths phase which is located on the sub-critical side of the
critical point is characterized by a line of fixed points where  
the rates scale as
\be
|\ln\tilde\lambda|\sim\xi, \qquad \tilde\mu\sim\xi^{-z}, 
\ee
while the mass and the length behave as 
\be
\tilde n\sim\ln\xi, \quad \tilde l\sim\xi. 
\ee
The dynamical exponent $z$ appearing here depends on the distribution of
rates of the original model \cite{igloi}.   
In the super-critical Griffiths phase $\lambda$ and $\mu$, as well as $n$ and $l$ are interchanged in the above relations.

\section{Weak asymmetry}
\label{weak}

Returning to the asymmetric contact process, we introduce
a control parameter which measures the magnitude of  
infection rates relative to that of recovery rates by 
\be
\Delta\equiv \overline{\ln\omega_i}-\overline{\ln\mu_i},
\ee
where the over-bar denotes an average over sites,
and a local asymmetry parameter as
\be
a_i\equiv\ln\lambda_i-\ln\kappa_i.
\label{a}
\ee
First, let us consider a weak, asymmetric perturbation of the symmetric model:
the parameters $a_i$ are uncorrelated random variables written in the form
\be
a_i\equiv I\alpha_i,
\ee 
where $I$ is an infinitesimally small global constant and $\alpha_i$ is
$O(1)$. 
Let us assume, furthermore, that the transformation is close to the 
fixed point, so that 
the approximate but asymptotically exact rules in Eqs. (\ref{mu_appr}) and 
(\ref{inf_appr}) can be used. 
Now, it is expedient to use the variables $(\omega_i,a_i)$ instead of 
$(\lambda_i,\kappa_i)$ in terms of which the old variables are expressed as 
$\lambda_i=\omega_i(1+e^{a_i})$ and $\kappa_i=\omega_i(1+e^{-a_i})$. 
In a cluster elimination step these variables transform (asymptotically) as 
\beqn 
&&\tilde a\simeq a_1+a_2  \\
&&\tilde\omega\simeq f(a_1,a_2)\frac{\omega_1\omega_2}{\mu_2},
\label{omega} \\ 
&&f(a_1,a_2)\equiv 1+\frac{e^{a_1}+e^{a_2}}{1+e^{a_1+a_2}},
\label{f}
\eeqn
while in a cluster merging step they still follow the rule given 
in Eq. (\ref{mu_appr}). 
Since the parameters $a_i$ are infinitesimal, we have $f(a_1,a_2)=2$ 
and the pair of variables $\omega_i$ and $\mu_i$
transforms autonomously (i.e. not influenced by the variables $a_i$), 
identically to the infection rate and recovery rate
of a symmetric contact process.  
As can be seen above, the asymmetry parameter $a_i$ transforms in the same way 
as the length $l_i$ of effective links given in Eq. (\ref{l}). 
The only difference is in that the original parameters $a_i$ have not 
necessarily the same sign, while we have $l_i=1$.  
The renormalized asymmetry parameter $\tilde a_i$ can thus be written as a 
sum of original variables $a_i$ with $\tilde l_i$ terms which are selected 
by the procedure independently of the parameters $a_i$.  

The above renormalization scheme allows for two different scenarios 
of the evolution of the
asymmetry parameters depending on the initial distribution of rates. 
There may be initial distributions biased to the left or to the right
for which the asymmetry parameter $\alpha_i$ averaged over sites 
tends to plus or minus infinity, respectively.  
Or it may tend to zero as the fixed point of the transformation is
approached. In the latter case, the system is statistically symmetric, 
although it is locally asymmetric. 
A sufficient condition for this is that the distribution of original
infection rates is invariant under the interchange of $\lambda$ and $\kappa$,
i.e. 
\be
P(\lambda,\kappa)\equiv P(\kappa,\lambda). 
\label{sym_dist}
\ee
This property will then be
preserved by the renormalization. 
It is, however,  not a necessary condition for the system being statistically
symmetric but owing to the approximative nature of the renormalization 
far from the asymptotical region and that even the homogeneous model is
non-integrable it is not possible to find a general condition 
in terms of the distribution of the initial rates. 

On the basis of the above considerations,  
if the initial distribution of infection rates is biased so that the mean
asymmetry is growing in magnitude, it must scale close to the fixed point
as the length: 
\be 
\overline{\tilde\alpha(\xi)}\sim \overline{\tilde l(\xi)}.
\ee
But if the system is initially statistically
symmetric, i.e. $\overline{\tilde\alpha(\xi)}\to 0$ as the fixed point is
approached, then, according to the central limit theorem, 
the variance of
$\tilde\alpha$ scales as the length:
\be 
\overline{\tilde\alpha^2(\xi)}\sim \overline{\tilde l(\xi)}.   \qquad
(\overline{\tilde\alpha}\to 0)
\ee
Using now the results of the symmetric model quoted in the previous section,
we obtain for the scaling of the typical
asymmetry parameter at the critical point $\Delta=\Delta_0$
\beqn 
|\tilde\alpha(\xi)|\sim \xi^{d_f/2}, \quad {\rm if} \quad \overline{\tilde\alpha}\to 0 \\
\tilde\alpha(\xi)\sim \xi^{d_f},\quad {\rm otherwise.}
\label{crit_d}
\eeqn
In the sub-critical Griffiths phase ($\Delta<\Delta_0$), we obtain
\beqn 
|\tilde\alpha(\xi)|\sim \xi^{1/2}, \quad {\rm if} \quad \overline{\tilde\alpha}\to 0 
\label{sub} \\
\tilde\alpha(\xi)\sim \xi,\quad {\rm otherwise,}
\label{sub_d}
\eeqn 
while in the super-critical Griffiths phase ($\Delta>\Delta_0$)
\beqn 
|\tilde\alpha(\xi)|\sim (\ln\xi)^{1/2}, \quad {\rm if} \quad
\overline{\tilde\alpha}\to 0 
\label{super} \\
\tilde\alpha(\xi)\sim\ln\xi,\quad {\rm otherwise.}
\label{super_d}  
\eeqn

\section{The statistically symmetric model}
\label{symmetric}

Next, we consider finite local asymmetry parameters but require that 
the distribution of $a_i$ is 
not too broad so that its moments are finite.  
As we have seen above, the asymmetry parameter scales 
differently  depending on whether the system is
statistically symmetric or not. 
Let us consider first, the former case, 
which is guaranteed if Eq. (\ref{sym_dist}) holds. 

Assume temporarily that the initial asymmetry is finite but weak 
($I\ll 1$) so that, at least within a finite length scale $\xi_0(I)$, 
the system can be regarded as approximately symmetric. 
As the model is renormalized, the initially uncorrelated rates $\lambda_i$ and
$\kappa_i$ become correlated and,  
if the system is critical, the typical asymmetry parameter starts to grow as 
$|\tilde a|\equiv|\ln\tilde\lambda-\ln\tilde\kappa|\sim I\xi^{d_f/2}$. 
But the typical value and the width of the distribution of logarithmic rates
themselves are increasing faster, see Eq. (\ref{symm_crit}), since $d_f/2<1/2$. 
Consequently, if a combined infection rate $\omega_i$ is 
much smaller than $\mu_2$ in a
cluster elimination step, then 
typically both $\lambda_i$ and $\kappa_i$ are much smaller than $\mu_2$. 
In other words, a complete separation of levels is realized. 
Furthermore, 
taking the logarithm of Eq. (\ref{omega}), we can see that the term 
$\ln f(a_1,a_2)$, which is at most $O(\xi^{d_f/2})$, is negligibly small 
compared to
the other terms on the r.h.s., which are $O(\xi^{1/2})$.
Therefore the SDRG transformation can be continued beyond
the scale $\xi_0(I)$ and as the fixed point is approached
($\xi\to\infty$), the relative asymmetry parameter tends to zero with
probability one, 
\be 
\frac{\ln(\tilde\lambda_i/\tilde\kappa_i)}{\ln\tilde\lambda_i}\sim\xi^{(d_f-1)/2}\to 0, \qquad (\Delta=\Delta_0)  
\label{rel_crit}
\ee
and the system transforms asymptotically like the symmetric one. 
But this is not restricted to the critical point. 
Following the above reasoning and using Eq. (\ref{sub}), we obtain a vanishing
relative asymmetry parameter in the sub-critical Griffiths phase,
\be 
\frac{\ln(\tilde\lambda_i/\tilde\kappa_i)}{\ln\tilde\lambda_i}\sim\xi^{-1/2}\to 0, \qquad   
 (\Delta<\Delta_0)
\label{rel_sub}
\ee
as well as, using Eq. (\ref{super}), in the super-critical Griffiths phase:
\be 
\frac{\ln(\tilde\lambda_i/\tilde\kappa_i)}{\ln\tilde\lambda_i}\sim(\ln\xi)^{-1/2}\to 0. \qquad   (\Delta>\Delta_0)
\label{rel_super}
\ee
So, we conclude that, in the statistically symmetric model, the local
asymmetry becomes irrelevant in the fixed point of the SDRG transformation, 
and, consequently, the large scale properties are identical to that of the symmetric model. 

It is, however, difficult to answer whether for any type of initial
distributions of transition rates the large scale properties of the system 
are described by the fixed point of the SDRG transformation. 
For contributions to the unresolved question whether any weak initial disorder 
results in logarithmic critical scaling described by the infinite-randomness
fixed point in the symmetric model we refer the reader to
Refs. \cite{hiv,vd,nft,hoyos}. 
In the asymmetric contact process this
problem is made more difficult by introducing the extra parameter $a_i$. 
We have numerically implemented the SDRG scheme and seen that, for not only 
weak but even for relatively strongly asymmetric initial disorder, 
such as independent uniform
distributions of the rates $\lambda_i$, $\kappa_i$ and $\mu_i$, 
the critical behavior is controlled by the infinite randomness fixed point. 
The numerical results (not shown) have been found to be in agreement with 
the analytical predictions about the
scaling of renormalized rates and asymmetry parameter both in the critical
point and in the Griffiths phases. 

\subsection{The site-symmetric model} 

We make a digression here on a special model studied 
by Monte Carlo simulations \cite{vd}, where 
the infection spreads from each site in both directions with equal rates, 
i.e. $\lambda_i=\kappa_{i-1}$ for all $i$. 
Note that this model is different from the symmetric one, where 
$\lambda_i=\kappa_i$ for all $i$.
As opposed to the earlier assumption, the initial asymmetry 
parameters are now correlated variables and it is easy to see that 
the sum of $a_i$ over subsequent sites is independent from the number of
terms: 
\be
\sum_{i=1}^na_i=\ln(\lambda_1/\kappa_n).
\label{corr}
\ee  
Thus, one expects here a slower increase of the asymmetry parameter under
renormalization than for the uncorrelated disorder. 
Assume for the sake of simplicity that the initial disorder is strong enough
so that the asymptotic forms of the renormalization rules can be used.  
The renormalized asymmetry parameter of an arbitrary effective link is 
the sum of initial variables
$a_i$ of links which have been incorporated into the effective link via
cluster eliminations. These links, when regarded in the original
system, form a disconnected set due to the eventual 
cluster merging steps occurring during the renormalization. 
Each connected part of length $n$ of this set gives a contribution given in
Eq. (\ref{corr}) independently of $n$ to $\tilde a$.  
So, the number of independent terms in $\tilde a$ is given just by the number
of connected components of the set of links rather than the total number of
links.   
It is easy to obtain that the number of connected components $c_i$, which is
initially $1$ on each link,
transforms in a cluster elimination step in the way: 
\be
\tilde c=\left\{
\begin{array}{c}
c_1+c_2-1, \quad {\rm if} \quad n_2=1 \\
c_1+c_2, \quad {\rm if} \quad n_2>1 \\
\end{array}
\right.
\ee
This is not much different from the transformation of the variable $l_i$
given in Eq. (\ref{l}) and, taking into account that the mass of clusters is
increasing, the probability of finding a cluster with $n=1$ goes to zero as
$\xi\to\infty$. So, the transformation rules of $n$ and $c$ are asymptotically
identical and, consequently, the scaling of the asymmetry parameter follows
the same laws as in the uncorrelated case given in
Eqs. (\ref{rel_crit}-\ref{rel_super}), and at most the prefactors are
reduced.

\section{The biased model}
\label{biased}

In the remaining part of the paper we will discuss the case 
when the spreading of activity is biased to either
direction, for example, if $a_i>0$ for all $i$, 
and will be interested in the possible 
fixed points of the SDRG transformation.
We have seen for infinitesimal asymmetry that
$\overline{\tilde\alpha}\to\infty$ in the biased model. 
For finite asymmetry, we will therefore have typically 
$\tilde\lambda_i\gg\tilde\kappa_i$ on large scales
and, consequently,  
$\tilde\omega_i\simeq\tilde\kappa_i$ and $f(\tilde a_1,\tilde a_2)\simeq 1$. 
This has also been confirmed by a numerical implementation of the SDRG scheme. 

\subsection{Griffiths phase}

Assume now that the control parameter $\Delta$ is small enough, 
such that the system is sub-critical and 
even the forward rate $\tilde\lambda$ is typically 
small compared to $\tilde\mu$, i.e. 
$\tilde\lambda/\tilde\mu\to 0$. 
In this case, we have a complete separation of levels
in cluster elimination steps and 
the variables $\omega$ and $\mu$ transform autonomously as 
\be 
\tilde\omega\simeq\frac{\omega_1\omega_2}{\mu_2}, \qquad 
\tilde\mu\simeq\frac{\mu_1\mu_2}{\omega_2}, 
\ee
in cluster elimination and merging steps, respectively, whereas the 
asymmetry parameter in the former as 
$\tilde a\simeq a_1+a_2$.
The renormalized parameters under these transformations will scale 
asymptotically as 
\beqn
\ln\tilde\omega^{-1}\simeq\ln\tilde\kappa^{-1}\sim \xi, \nonumber \\
\tilde\mu\sim \xi^{-z}, \nonumber \\
\tilde a\sim\xi,
\eeqn
and flow to a line of fixed points parameterized by a 
non-universal dynamical exponent 
$z$ that depends on the distribution of initial rates \cite{igloi}.   
This line can be interpreted as the subcritical Griffiths
phase, where the dynamics are described by power laws with 
distribution-dependent exponents.

As in the Griffiths phase of the symmetric model, a finite length scale
$\xi_1$ above which the renormalized recovery rates are
typically larger than the renormalized combined infection rates can be defined.
This length scale can be interpreted as the characteristic size of locally
super-critical clusters. Beyond this scale, $\xi\gg\xi_1$, practically no
cluster merging step occurs in the SDRG procedure. 
In the biased model at this length scale $\xi\sim\xi_1$, the backward 
rates $\tilde\kappa$ are comparable with $\tilde\mu$ but the forward rates 
$\tilde\lambda$ are still larger than $\tilde\mu$. 
Similar to $\xi_1$, one can define a second length scale $\xi_2(>\xi_1)$
above which the recovery rates typically exceed even the forward rates. 
An interpretation of $\xi_2$ can be given if the process starts from a single
active site. In this case, the activity spreads to the right typically up to a
finite distance $\xi_2$, where it is trapped until becoming extinct. 
Thus, the observables usually measured in simulations, such as the survival
probability, are expected to have the same time dependence beyond a time
scale corresponding to $\xi_2$ as in the sub-critical Griffiths phase of the
symmetric model.

\subsection{The lower phase transition and the weak-survival phase}

When the control parameter $\Delta$ is increased,
both $\xi_1$ and $\xi_2$ increase and there must be a point $\Delta=\Delta_1$,
where $\xi_2$ diverges but $\xi_1$ is still finite. 
This point corresponds to a transition to a phase where the 
activity spreads to the right without limits in a finite fraction of 
random environments, 
and the survival probability tends to a positive limit as $t\to\infty$. 
At this point and above, $\tilde\mu$ will not exceed $\tilde\lambda$ 
asymptotically, therefore there is only a partial separation of levels.
As aforementioned, we obtain then no correct information on how 
the activity spreads rightwards by the SDRG method and it is therefore unable 
to predict the properties of this phase transition. 
The fact that the level corresponding
to $\lambda$ does not separate suggests that the transition is, at least, 
not of activated type characterized by logarithmic time-dependence of observables. Nevertheless, the lowest level separates well from the other ones even if
$\Delta\ge \Delta_1$ and in the asymptotic
transformation of the corresponding rate $\kappa$ and that of $\mu$ the 
'incorrect' rate $\lambda$ does not play a role.   

Although we cannot infer the properties of the phase transition, the 
spreading of the front of the activity started from a single active seed
can be related to the distribution of the renormalized recovery rates.
If the system is renormalized well beyond the scale $\xi_1$, the renormalized
backward rates are negligibly small and it behaves as
a totally asymmetric contact process, where the recovery rates have a broad
distribution with an algebraic tail
\be 
P_<(\mu)\sim\mu^{1/z}, \quad \mu\to 0,
\label{mu_dist}
\ee 
as can be inferred from the form of the fixed point distribution of $\mu$
given in Ref. \cite{igloi}. 
Assume now that the process was started from a 
single active seed at $x(t=0)=0$ and at time $t$ there are still active sites.
If the leftmost active site, the position of which is 
denoted by $x(t)$, recovers, it will be practically not 
reinfected and $x$ is shifted to the closest active site on its right. 
The expected value of
the time of recovery is $1/\mu_i$, therefore the mean time $t_{1,n+1}$ 
needed for $x$ to
shift from site $1$ to site $n+1$ is at most the sum of recovery times: 
\be
t_{1,n+1}<\sum_{i=1}^{n}\mu_i^{-1}. 
\ee
For large $n$, the asymptotical behavior of this sum is governed by 
the exponent $z$. 
If $z<1$, $\sum_{i=1}^{n}\mu_i^{-1}\sim n$, whereas if $z>1$, 
$\sum_{i=1}^{n}\mu_i^{-1}\sim n^z$. 
This yields the following lower bound on the asymptotical displacement of the
leftmost active site:
\beqn 
x(t)>O(t)\quad {\rm if} \quad z<1 \nonumber \\
x(t)>O(t^{1/z})\quad {\rm if} \quad z>1.
\label{bound}
\eeqn   
Obviously, the position of the rightmost active site 
(the front of the activity)
must move at least as fast as $x(t)$ in samples which are surviving
up to time $t$. 
If $\Delta>\Delta_1$, the forward infection rates are typically larger than 
the recovery rates in the renormalized model, therefore the density of active
sites between $x(t)$ and the front is expected to be finite, similar to the
super-critical phase of the symmetric model. 
Consequently, if the leftmost active site recovers, the expected 
shift of $x(t)$ is finite and the relation in (\ref{bound}) holds as a
proportionality.  

So, the active cluster moves at least as a power of time, which is
different from the behavior of the unbiased model.
If $z>1$, which is realized by a weak initial bias,  
the above reasoning suggests the creeping motion of the
critical active cluster with an asymptotically vanishing velocity. 
This is in contrast to the homogeneous, biased model 
where the active cluster moves with a finite velocity.  

\subsection{The upper phase transition and the strong-survival phase}

If the control parameter $\Delta$ is further increased in the
weak-survival phase ($\Delta>\Delta_1$), the length scale $\xi_1$, 
as well as the 
dynamical exponent $z$ increase and, at a second phase transition point 
$\Delta=\Delta_2$, they diverge. 
Physically, this means that for $\Delta>\Delta_2$, the activity, 
when started from a single active seed spreads also leftwards without limits. 
In the transition point, the variables $\omega\simeq\kappa$ and $\mu$ are
dual to each other and the fixed point of the SDRG transformation is an
infinite randomness fixed point with the asymptotic scaling of parameters: 
\be 
|\ln\tilde\kappa|\sim |\ln\tilde\mu |\sim\xi^{1/2}. \qquad (\Delta=\Delta_2)\label{delta2}     
\ee
Whereas, above $\Delta_2$, we obtain a line of fixed points 
parameterized by a distribution-dependent dynamical exponent $z'$, 
where the parameters scale as   
\be
\tilde\kappa\sim \xi^{-z'},  \qquad |\ln\tilde\mu |\sim\xi.  \qquad (\Delta>\Delta_2)
\label{s_surv}
\ee 

The phase transition at $\Delta=\Delta_2$ can be also detected in the change 
of the finite-size scaling  
of the lowest non-trivial eigenvalue of $Q$ in a finite, 
open system of size $L$, which is (asymptotically) correctly 
treated by the SDRG method.  
The inverse of this quantity that gives the time scale of 
reaching the absorbing state from the fully active one scales 
in the following  way: 
\beqn
\tau\sim L^{\max(1,z)}, & \quad \Delta_1<\Delta<\Delta_2 \nonumber \\  
\ln\tau\sim L^{1/2},& \quad \Delta=\Delta_2 \nonumber \\
\ln\tau\sim L. &\quad   \Delta>\Delta_2  
\label{fss}
\eeqn
Note that in a finite but periodic system, the finite-size scaling of 
the lowest gap of $Q$ is different from this. In that case, the extinction time
behaves as $\ln\tau\sim L$ in the entire range $\Delta>\Delta_1$, since 
the activity can 'go around' rightwards and it is irrelevant whether
the spreading of infection leftwards is blocked or not. 
This is consistent with the SDRG treatment. When the periodic system is
renormalized up to two effective sites then in the last decimation step 
the level which was the lowest (non-trivial) one till that point and 
which could indicate the phase transition at $\Delta_2$ is lost. 

As we mentioned above, in the biased model, the renormalized forward rates will be much greater than the backward ones.  
So, it is plausible to assume that the behavior of the system 
in the range $\Delta>\Delta_1$
is well described by a simply tractable idealized model where 
the forward rates are infinitely large $\lambda_i=\infty$. 
If this process is started from a single active site, the activity will immediately
spread to the right without limits, and all sites on the right hand side of
the leftmost active site will be active since, once a site recovers, its left
hand side neighbor will instantly reinfect it. 
It is easy to see that the position $x(t)$ of the leftmost active site is
a random walk with jump rate $p_i=\mu_i$ and $q_i=\kappa_{i-1}$ to the right and 
to the left, respectively if $x(t)=i$. 
The basic properties of the one-dimensional random walk in a 
random environment are exactly
known, for a review see e.g. Ref. \cite{bouchaud}. 
Varying the control parameter of the problem,  
\be
\Delta_{\rm RW}=\overline{\ln q_i}-\overline{\ln p_i},
\ee
the expected displacement $x(t)$ in typical environments has 
different asymptotics:
\beqn 
x(t)\sim t^{1/z}  \qquad  \Delta_{\rm RW}<0 \nonumber \\
|x(t)|\sim (\ln t)^2  \qquad  \Delta_{\rm RW}=0 \nonumber \\
-x(t)\sim t^{1/z}  \qquad  \Delta_{\rm RW}>0,
\eeqn 
where $z$ is the positive root of the equation
$\overline{(q_i/p_i)^{1/z}}=1$ for $\Delta_{\rm RW}<0$.

So, in the range  $\Delta_1<\Delta<\Delta_2$, the position of the leftmost active site is expected to creep rightwards, at 
$\Delta=\Delta_2$ to perform a Sinai walk \cite{sinai} 
with a zero average displacement
and fluctuations in the order of $(\ln t)^2$, finally, if $\Delta>\Delta_2$, to
creep leftwards.    
These results are consistent with the finite-size scaling of the 
extinction time given in Eq. (\ref{fss}) and 
are in line with the behavior of the homogeneous, biased
contact process, where $x(t)$ is a random walk in a homogeneous environment 
and the extinction time scales at the second transition point 
as $\tau\sim L^2$ \cite{2nd}.

The phase transition at $\Delta=\Delta_2$ can also be detected  
in the behavior of the survival probability in a semi-infinite system, 
when the process is started from a single active site (site $0$) and infection
spreading from this site is not possible to the right, i.e. $\lambda_0=0$. 
This quantity can be calculated in the idealized model, where  
the site on the right hand side of the origin of the random walk will be an
absorbing site.      
The probability $P(x)$ averaged over the random environments (i.e. sets of
transition rates) that the walker visits a site in a distance $x$ from the
origin before it hits the absorbing site is known to be 
$O(e^{-{\rm const}\cdot x})$,
$O(x^{-1/2})$ and constant for $\Delta_{\rm RW}<0$, $\Delta_{\rm RW}=0$ and 
$\Delta_{\rm RW}>0$, respectively, see e.g. Ref. \cite{ir}. 
Using these results, the asymptotical time-dependence of the survival
probability can be derived. 
If $\Delta<\Delta_2$, the activity is localized in the vicinity of the origin
but, due to the algebraic tail of the distribution of 
effective recovery rates in Eq. (\ref{mu_dist}), 
the survival probability averaged over disorder decays algebraically, as well: 
\be 
P_{\rm surv}(t)\sim t^{-1/z}. \qquad (\Delta<\Delta_2)
\ee
At the second critical point, $\Delta=\Delta_2$, using the relation between
time and length scale of the Sinai walk $\ln t\sim \xi^{1/2}$, which can be
gathered also from the fixed point solution of the SDRG transformation 
in Eq. (\ref{delta2}), we obtain for the survival probability: 
\be    
P_{\rm surv}(t)\sim(\ln t)^{-1}. \qquad (\Delta=\Delta_2)
\ee
This form is identical to the surface critical behavior 
of $P_{\rm surv}(t)$ in the unbiased model \cite{hiv}. 
An important difference is, however, that the set of active sites 
is compact in the biased model, i.e. has a finite density, while in
the symmetric model it is a fractal of dimension $d_f$. 

Finally, for $\Delta>\Delta_2$, the survival probability tends to a 
positive limit that depends on the initial distribution of rates when $t\to\infty$.

\section{Summary and outlook} 
\label{summary}

We have applied an SDRG method to the asymmetric contact process where the
infection rates to the left and to the right, as well as the recovery rates 
are independent random variables. 
We have shown that the local random asymmetry in the infection spreading is
irrelevant on large scales if the model is globally (statistically) symmetric,
and the critical and off-critical behavior is identical to that of the
disordered, locally symmetric contact process. 
If the model is  globally biased, the SDRG transformation 
predicts two distinct phase transitions. 
The lower one is related to the transmission of infection in the
direction of the bias. 
Unlike this transition point, which is out of the range of validity of the
method, the upper critical point, which is related to the spreading of
activity against the direction of the bias, 
is controlled by an infinite randomness fixed point. 

It would be desirable to check the predictions of the SDRG method by Monte
Carlo simulations. An intriguing question which is left open by the 
SDRG analysis is
nature of the lower phase transition. This is predicted to be neither of
activated type with logarithmic critical scaling nor to be in the directed
percolation class at least for weak asymmetry when the critical cluster is
creeping. 
Preliminary results of Monte Carlo simulations (not shown) confirm this 
but, to clarify the properties of the phase transition, 
an extensive numerical analysis would be needed. 
The present work is restricted to $d=1$, although 
the critical behavior of the symmetric contact process 
in the presence of disorder
is known to be governed by an infinite randomness fixed point also in dimensions
$d>1$ \cite{kovacs,monthus,vfm}. 
The question of stability of the fixed point against asymmetry in higher
dimensions remains a challenging open problem.

\acknowledgments
This work was supported by the J\'anos Bolyai Research Scholarship of the
Hungarian Academy of Sciences, by the National Research Fund 
under grant no. K75324, and partially supported by the European Union and the
European Social Fund through project FuturICT.hu (grant no.:
TAMOP-4.2.2.C-11/1/KONV-2012-0013).

\appendix
\section{The spectrum of the rate matrix}

\subsection{Two-site block}

The rate matrix of a two-site block shown in Fig. \ref{rg} is given as: 
\be 
Q_{12}=
\begin{pmatrix}
0 &0 & 0& 0\\
\mu_1 &-\mu_1-\lambda_1 &0 &\lambda_1\\
\mu_2 &0 &-\mu_2-\kappa_1 &\kappa_1 \\
0 &\mu_2 &\mu_1 &-\mu_1-\mu_2
\end{pmatrix}.
\ee
The three non-trivial eigenvalues are roots of the 
cubic equation:
\be 
\epsilon^3+a\epsilon^2+b\epsilon+c=0,
\label{cubic}
\ee
where 
\beqn
&&a=\lambda_1+\kappa_1+2(\mu_1+\mu_2) \nonumber \\
&&b=\lambda_1\kappa_1+\mu_1\mu_2+(\mu_1+\mu_2)(\mu_1+\mu_2+\lambda_1+\kappa_1)
\nonumber \\
&&c=\mu_1\mu_2(\mu_1+\mu_2+\lambda_1+\kappa_1).
\eeqn

\subsection{Three-site block}

The rate matrix of the three-site block shown in Fig. \ref{rg2} has a block diagonal form: 
\begin{widetext}
\be 
Q_{123}=
\begin{pmatrix}
0 & 0& 0 & 0& 0& 0& 0& 0 \\
\mu_2 & -\kappa_1-\lambda_2-\mu_2 & 0& \kappa_1& 0 &\lambda_2 &0 &0 \\
0& 0& -\lambda_1& \lambda_1 & 0& 0& 0& 0 \\
0& 0& \mu_2& -\lambda_2-\mu_2 & 0& 0& 0& \lambda_2 \\
0& 0& 0& 0& -\kappa_2 &\kappa_2 & 0& 0\\
0& 0& 0& 0& \mu_2& -\kappa_1-\mu_2& 0 &\kappa_1 \\
0& 0& 0& 0& 0& 0& -\lambda_1-\kappa_2& \lambda_1+\kappa_2 \\
0& 0& 0& 0& 0& 0& \mu_2& -\mu_2 
\end{pmatrix}
\ee
\end{widetext}
and has the following eigenvalues: 
\beqn
&&\epsilon_1=\epsilon_2=0, \\
&&|\epsilon_{3,4}|=\frac{1}{2}(\mu_2+\lambda_1+\lambda_2)\times \nonumber \\
&&\qquad \times\left(1\mp\sqrt{1-4\lambda_1\lambda_2/(\mu_2+\lambda_1+\lambda_2)^2}\right)
\\
&&|\epsilon_{5,6}|=\frac{1}{2}(\mu_2+\kappa_1+\kappa_2)\times \nonumber \\
&&\qquad \times\left(1\mp\sqrt{1-4\kappa_1\kappa_2/(\mu_2+\kappa_1+\kappa_2)^2}\right)
\\
&&|\epsilon_7|=\mu_2+\lambda_1+\kappa_2, \quad |\epsilon_8|=\mu_2+\lambda_2+\kappa_1. 
\eeqn


\end{document}